\documentstyle[12pt,aasms4]{article}

\begin{document}

\def\be{\begin{equation}}
\def\ee{\end{equation}}
\def\bea{\begin{eqnarray}}
\def\eea{\end{eqnarray}}
\def\c{\cite}
\def\nn{\nonumber}

\def\et{ {\it et al.}}
\def\la{ \langle}
\def\ra{ \rangle}
\def\ov{ \over}
\def\ep{ \epsilon}

\def\mdot{\ifmmode \dot M \else $\dot M$\fi}    
\def\mxd{\ifmmode \dot {M}_{x} \else $\dot {M}_{x}$\fi}
\def\med{\ifmmode \dot {M}_{Edd} \else $\dot {M}_{Edd}$\fi}
\def\bff{\ifmmode B_{f} \else $B_{f}$\fi}

\def\apj{\ifmmode ApJ \else ApJ \fi}    
\def\apjl{\ifmmode  ApJ \else ApJ \fi}    %
\def\aap{\ifmmode A\&A \else A\&A\fi}    %
\def\mnras{\ifmmode MNRAS \else MNRAS \fi}    %
\def\nat{\ifmmode Nat\else Nat\fi}
\def\prl{\ifmmode Phys. Rev. Lett. \else Phys. Rev. Lett.\fi}
\def\prd{\ifmmode Phys. Rev. D. \else Phys. Rev. D.\fi}

\def\ms{\ifmmode M_{\odot} \else $M_{\odot}$\fi}    
\def\no{\ifmmode \nu_{1} \else $\nu_{1}$\fi}    
\def\nt{\ifmmode \nu_{2} \else $\nu_{2}$\fi}    
\def\ntmax{\ifmmode \nu_{2max} \else $\nu_{2max}$\fi}    
\def\nomax{\ifmmode \nu_{1max} \else $\nu_{1max}$\fi}    
\def\nh{\ifmmode \nu_{\rm HBO} \else $\nu_{\rm HBO}$\fi}    
\def\nqpo{\ifmmode \nu_{QPO} \else $\nu_{QPO}$\fi}    
\def\nz{\ifmmode \nu_{o} \else $\nu_{o}$\fi}    
\def\nht{\ifmmode \nu_{H2} \else $\nu_{H2}$\fi}    
\def\ns{\ifmmode \nu_{s} \else $\nu_{s}$\fi}    
\def\nb{\ifmmode \nu_{burst} \else $\nu_{burst}$\fi}    
\def\nkm{\ifmmode \nu_{km} \else $\nu_{km}$\fi}    
\def\dn{\ifmmode \Delta\nu \else $\Delta\nu$\fi}    

\def\rs{\ifmmode R_{s} \else $R_{s}$\fi}    
\def\rmm{\ifmmode R_{M} \else $R_{M}$\fi}    
\def\rco{\ifmmode R_{co} \else $R_{co}$\fi}    
\def\rim{\ifmmode R_{ISCO} \else $R_{ISCO}$\fi}    

\title{On The Low Frequency Quasi Periodic Oscillations of  X-ray Sources}

\author{ C.M. Zhang }

\affil{National Astronomical Observatories,
Chinese Academy of Sciences, Beijing-100012, China\\
zhangcm@bao.ac.cn}

\vskip 5cm

ASTR1245---2nd revised---July 2004
\\
ASTR1245---3rd revised---Sept 2004
\\

\begin{abstract}
Based on the interpretation of the twin kilohertz Quasi Periodic
Oscillations (kHz QPOs) of X-ray spectra of  Low Mass X-Ray Binaries
 (LMXBs) ascribed
 to the Keplerian and the periastron precession frequencies
 at the inner disk respectively,
 we ascribe the low frequency(0.1-10 Hz) Quasi Periodic Oscillations (LFQPO)
and  HBO (15-60 Hz QPO for Z sources or Atoll sources) to the periastron
precession at some outer  disk radius. It is assumed
 that both radii are correlated  by
a scaling factor of 0.4.
 The conclusions obtained include: All QPO frequencies increase
with increasing
accretion rate. The theoretical relations between HBO
(LFQPO) frequency
and the  kHz QPO frequencies  are  similar to the measured
empirical formula.
\end{abstract}


\section{Introduction}

The QPO mechanisms of low mass X-ray binary (LMXB)
 have been paid much attention
 since  early
1996 (van der Klis  2000) following the discovery of
 kHz QPOs by  the Rossi X-ray Timing Explorer (RXTE).
The Z sources (Atoll sources), which are high (less)
luminous neutron star LMXB (Hasinger \& van der Klis 1989),
typically show four
distinct types of QPOs (van der Klis  2000).
These are the  normal branch
oscillation (NBO) $\simeq 5-20$~Hz, the
horizontal
branch oscillation (HBO) $\nh \simeq 15-60$~Hz, and the
kHz QPOs $\nt(\no) \simeq 200-1200$~Hz
that typically occur in
pairs with upper frequency $\nt$ and lower frequency $\no$.
In several Atoll sources, nearly
coherent burst oscillations $\nb \simeq 300-600$~Hz
 have also
been detected during thermonuclear Type~I X-ray bursts.
 These are thought to be at the spin frequency of the
neutron star (NS)
or double it (see e.g.,  Strohmayer et al. 1996,
Strohmayer \& Bildsten 2000, Muno 2004).
Moreover, the low-frequency (0.1 - 10 Hz) QPOs that
 have been detected
 in the accreting X-ray black hole candidate (BHC),
  and the QPO properties
of BHCs show many similarities with those
of NS  sources  and the QPO frequencies
of NS  and BHC follow a tight
and systematic
correlation over three orders of magnitude in frequency
 (Psaltis et al. 1999, Belloni et al. 2002).
  Nevertheless, the frequencies
$\nt$ and $\no$,
 as well as  $\nt$ and $\nh$,  follows very
similar relations in five Z sources (Psaltis et al. 1998).

However the frequency separation
between the upper and the lower kHz QPOs $\dn \equiv \nt - \no$
 is not a constant
lower or higher than  the inferred NS spin frequency,
but in some cases  decreases systematically
with instantaneous \mdot{}, e.g.
Sco X-1 (van der Klis et al. 1997), 4U1608$-$52
(Mendez et al. 1998a,b),
4U1735-44 (Ford et al. 1998)
 and
4U1728$-$34 (Mendez and van der Klis 1999). In
  the latter,  the observed
coherent burst frequency 364 Hz is higher than its  maximum
$\dn \sim $ 355 Hz (Mendez and van der Klis 1999).

Various theoretical models have been
proposed to account for the  QPO phenomena  in
X-ray binaries (for a review see e.g., Psaltis 2000).
In the early detections of the RXTE, the
upper kHz QPO $\nt$ was simply considered to originate
 from the Keplerian orbital frequency  at the preferred  radius
close to the compact object that experiences
the inner accretion flows, and the lower kHz QPO
 $\no$ was  attributed to
the beating  of this  frequency with the stellar spin
frequency $\ns$
 (Miller et al. 1998). However recently,   general
relativistic effects have been  invoked
 by Stella and Vietri (1999, hereafter SV99)
to account for kHz QPOs, and these
 can satisfactorily explain the varing  kHz QPOs separation $\dn$.
 In this  model of SV99, the twin kHz
QPOs are ascribed to the
 Keplerian frequency $\nu_{K}$
and the periastron frequency of material orbiting the central mass at some
disk radius, i.e.,
$\nu_{2} = \nu_K = (M/4\pi^2 r^3)^{1/2}$
 and $\nu_{1} = \nu_{K}[1 -  (1 - 6M/r)^{1/2}]\;$, where r is the
Schwarzschild coordinate distance and M is the gravitational mass
of the source (We set  the speed of light and the
gravitational constant $c=G=1$ throughout this paper).
 It is commonly believed that the kHz QPOs
 are produced  close to the innermost stable circular
orbit (ISCO) or the surface of  NS, which  provides  a probe
to detect
the accretion flow around the non-Newtonian strong gravity region
(van der Klis 2000, SV99).

 A theory of epicyclic parametric resonance
 in a relativistic accretion disk
was proposed (Abramowicz \& Kluzniak 2001,
 Abramowicz et al. 2003), in which  the twin kHz  QPOs occur
 at the frequency of meridional
oscillation and the radial epicyclic frequency in the same orbit.
This can explain
the frequency ratio 3:2 detected
in black hole candidates (e.g., XTE J1550-564, Remillard et al. 2002).
Although many other viable new ideas have also been  proposed
(e.g.,  Wagoner 1999, Psaltis \& Norman 2000,
 Osherovich \& Titarchuk 1999,  Titarchuk et al. 1998, Klein et al.
1996),
there has not yet been any  model satisfactorily  explains
 all observed  QPO  phenomena.
The QPO mechanisms of accreting X-ray binaries
 are still  debated and there remain open problems. In the 1980's,
 HBO frequency ($\nh \simeq 15 - 60 $Hz)
 was interpreted as  the beat frequency between
the Keplerian frequency of the magnetosphere-disk  and the stellar
spin frequency using  the standard beat frequency model
(BFM) (Alpar and Shaham 1985, Lamb et al. 1985).
Later, this  was considered to be
 the nodal precession  due to the  Lense-Thirring
effect in the disk (Vietri and Stella 1998).
 Further, in the model of Titarchuk and Wood (2002),
 HBO (as a low  frequency $\nu_{low}$) was ascribed to the
 magnetoacoustic oscillation frequency, and then  a linear
correlation between this low frequency and the high frequency
$\nu_{high}$ (identified to be the Keplerian frequency)
 is inferred.

In this article,  we concentrate on the explanation of
 LFQPO ($\nu_{low}$) or HBO of luminous  NS X-ray sources
 (for Atoll sources,  15 - 60 Hz QPO is supposed to be
 due to the
 same mechanism as HBO of Z sources, see e.g., Psaltis et al. 1999),
and their relationship  to the twin kHz QPOs. However  we neglect
the details of the physical mechanism for QPO production
at the present time. A summary and conclusions are given
in the final section.

\section{The Model and its Results}
The accretion flow of the matter around NS and BH is
complicated, especially close to the
innermost stable circular orbit (ISCO), the radius
$\rim = 3\rs$, where $\rs = 2M \sim 3m$ (km) (one solar mass
corresponds to about 3 km of the Schwarzschild radius)
 is the Schwarzschild radius calculated using   the gravitational mass
 M, where $m = M/\ms$ is the mass in units of solar mass.
In reality, the motion of the accreted  matter will be
influenced not only by the gravitational field but also by the
 magnetic field. For simplicity, we assume that the disk matter
 that exhibits
QPO is mainly dominated by the Schwarzschild  gravitational field with
 a slightly eccentric orbit ($e\simeq 0$).

Here we assume  that HBO frequency $\nh$, as well as LFQPO $\nqpo$,
  is the  periastron  precession frequency
of the accreted orbiting  material
 at some outer disk  radius, and the twin kHz
QPOs are ascribed to the Keplerian frequency
and the periastron frequency  of the orbiting  material  at
some inner disk radius (as described by SV99).
 It is also assumed
that there exists a scaling factor $\phi$ to
connect the two radii, which
will be determined later  by the comparison of the QPO data
with the model.
Therefore, these QPO frequencies
are conveniently arranged as follows
 by  defining the parameter
$y\equiv {R_{s} \over r}$, which is the
 ratio of the Schwarzschild radius to the instantaneous inner  disk radius.

\be
\no (y) = (1-\sqrt{1-3y})\times {\; }  \nt (y)\;,
\label{no}
\ee
\be
\nt (y) = {\nz}  y^{3/2}\;,
\label{nt}
\ee

\be
\nh (y) =  \no (\phi y)\\
        =   {\nz} (1-\sqrt{1-3\phi y})\times    (\phi y)^{3/2}{\; }
\;,\;\;\;
\label{nh}
\ee

\be
\nz \equiv {11300\over m}{\;}{(Hz)}\;,
\label{phi}
\ee
where  $\phi$ is the  scaling parameter  connecting  the radii of the
 inner disk  and
 the outer disk,  which   we set   to be $0.3 \sim 0.4 $
  to give the best consistency  when comparing the model
with the observational  data.

From Eq.(\ref{no})  and Eq.(\ref{nt}),
 the parameter y can be expressed as,
\be y= {1\ov 3}({\no \ov \nt})[2 - {\no \ov \nt}]\;,\;\;\; y = 0.2
({m\nt \ov 1000})^{2/3}\;. \ee
 Considering a  maximum value
y=1/3 at ISCO from Eq.(\ref{no}),
 the parameter y can  be approximately acquired using
 a Taylor expansion of $\phi y$ to the first order in
Eq.(\ref{nh}) since  the maximum value of $\phi y \sim 0.4/3 \sim
0.13$, hence

\be
y \simeq \phi^{-1} ({2\nh \ov 3\nz})^{2/5}\;.
\ee
At the innermost stable circular orbit (ISCO), the parameter
$y={1\ov 3}$ and the maximum twin kHz QPO frequencies are given by
$\nu_{Max}=\no(1/3)=\nt (1/3)={2175\over m} (Hz)$, corresponding  to 1200 (Hz)
 for  1.9 solar  masses.  The mass of the source can also be expressed by
the maximum kHz QPO frequency at ISCO,
\be
m=\frac{2175}{\nu_{Max}}\;.
\label{massmax}
\ee

The relation  $\nh$ vs.  $\nt$ is plotted in
FIG.1, together with the  measured five Z-source samples,
and it is seen  that the agreement  between the model and
the observed QPO
data is quite good  for the   selected values of the NS
 mass of about m=2.0 ($\ms$) and  of $\phi=0.4$, which are the two
  free parameters in the Eqs.(1-3).

In FIG.2,
we take the low-frequency QPO  $\nqpo$ of Atoll-sources, as well as
several other NS systems and a number of BH
 binaries for a  similar mechanism to HBO
of the Z-sources
(Psaltis et al. 1999),
and plot the theoretical curve with respect to lower kHz QPO $\no$.
 We find  remarkably consistent correlations between
the theoretical expectations and the detected data
which extends over nearly three orders of magnitude in  QPO
frequency. However one interesting fact is that the theoretical curves
are only weakly related to the stellar masses because
 $\no$ and $\nqpo$ are all inversely
related to the  mass parameter in  Eqs.(\ref{no}) and (\ref{nh}). Their
ratio  $\no/\nqpo = \phi^{-3/2} (1-\sqrt{1-3 y})/(1-\sqrt{1-3\phi y})$
has no direct correlation with stellar mass. Further we have
 for $\phi=0.4$, $1\le LOG(\no)-LOG(\nqpo)\le 1.25$, which means that
the theoretical curves show  little variation
 in the $\nu_{QPO}$ vs. $\no$ diagram  as a log  scale even if the mass
changes by one order  of magnitude. We plot  theoretical curves with
parameter conditions
 $m=5.0 \ms$ and $\phi=0.4$ (short dashed line), as well as
 $m=10.0 \ms$ and $\phi=0.4$ (long dashed line), and find that
  the theoretical curves are insensitive to the mass parameter.
 However,  the mass parameter will influence on the maximum frequency
of the lower kHz QPO at ISCO because it is inversely related to the mass
from Eq.(\ref{massmax}).
 For example, if 300 Hz QPO in GR 1655-40 were the maximum frequency,
 then the  mass of this source would be 7.0 solar masses, inferred  from
Eq.(\ref{massmax}).  It is
also seen that $\nu_{\rm HBO}$ increases slowly  with
$\nu_1$  when $\nu_1$ is higher, as anticipated  by Psaltis et al. 1999.

The weak correlation with source mass   in the
$\nu_{QPO}$ vs. $\no$ diagram  seems to hint that the QPO phenomena of
Atoll sources, Z-sources, as well as other
 accreting X-ray binaries,
 are  intimately related to the specific radii of the
disk and the
ratio between them. This  may reflect  a  common property of the
accretion flow around the gravitational sources. However the mechanism
to account for this is still unclear.

 From Eqs.(\ref{no}), (\ref{nt}) and (\ref{nh}),
we can derive the theoretical
relations between QPO frequencies   which are the
following,

\be
\nh \simeq 500 \phi^{5/2}{\;}(Hz){\;}({\no\over 500})
[ 1 - 0.19 (1-\phi) ({m\no \over 500})^{2/5} ] \;,
\label{nhno}
\ee

\be
\nh \simeq 297.3 \phi^{5/2} \;(Hz)\;m^{2/3}
({\nt \over 1000})^{5/3}[1 + 0.15 \phi ({m\nt\over 1000})^{2/3}] \;,
\label{nhnt}
\ee

\be
\no \simeq 297.3{\;}(Hz){\;}m^{2/3}({\nt\over 1000})^{5/3}
[1 + 0.15({m\nt\over 1000})^{2/3}] \;.
\label{nont}
\ee
If we set the orbit scaling factor $\phi=0.4$, we have the following
 specific relations,

\be
\nh \simeq 50.6 {\;}(Hz){\;}({\no\over 500})
[ 1 - 0.11 ({m\no \over 500})^{2/5} ] \;,
\label{nhno2}
\ee

\be
\nh \simeq 30.1 {\;}(Hz){\;}m^{2/3}({\nt\over 1000})^{5/3}
[1 + 0.06({m\nt\over 1000})^{2/3}] \;.
\label{nhnt2}
\ee
As a comparison, we list the empirical
relations between HBO and twin kHz QPO frequencies
 (Psaltis et al. 1998, Psaltis et al. 1999),

\be
\nu_{\rm HBO}\simeq (42\pm3~Hz)(\nu_1/500~Hz)^{0.95\pm0.16}\;,
\label{nhnoe}
\ee

\begin{equation}
 \nu_{\rm HBO} = 13.2\, a_2
 \left(\frac{\nu_{s}}{300~\mbox{Hz}}\right)
 \left(\frac{\nu_2}{1~\mbox{kHz}}\right)^{b_2}\;,
\label{nhnte}
 \end{equation}
 with $a_2 \approx 4.6$, and $b_2 \approx 1.8$. Therefore a
 consistency
between  Eq.(\ref{nhno2}) and  Eq.(\ref{nhnoe}), as well as
 between Eq.(\ref{nhnt2}) and  Eq.(\ref{nhnte}), is apparently found.
 Furthermore, we can  also obtain the NS mass formula represented  by
the twin kHz QPO frequencies from Eqs.(1-2),

\be
m =  2.17 \times ({\no \over 500})^{3/2} ({\nt\over 1000})^{-5/2}
[1 - {\no \ov 2\nt}]^{3/2}\; .
\label{mot}
\ee
According to Eq.(\ref{mot}), we can apply the detected twin
kHz QPO data (van der Klis 2000)  to calculate the stellar mass
 and obtain the mass average value 1.86 $\ms$ for the six Z-sources and
eleven Atoll sources, which is close to the measured
NS mass $1.78\pm0.23{}{} (\ms)$ of  X-ray binary Cygnus X-2
 (Orosz \& Kuulkers 1999).


\section{Summary  and Conclusions}

We have derived a  theoretical relation between HBO and $\no$,
 as well as HBO and $\nt$, and found that the detected low-high frequency
correlation (Psaltis et al. 1999, Belloni et al. 2002) can be
interpreted  if $\nu_{low}$ and $\nu_{high}$ are identified
as HBO frequency and $\no$ respectively.
 Furthermore, from Eq.(\ref{nhno2}),
if we set the parameters m=1.8 $\ms$ and $\no=700$ Hz, then we have
the approximate relation $\nh \simeq 0.08 \no$, which is very
close to the detected empirical relation given  in
Eq.(\ref{nhnoe}).  The correlation between $\nh$ and $\no$
  depends weakly  on the properties of sources, such as mass,
magnetic field and  the hard
surface of compact objects, which has been
 anticipated  by the observations (Belloni et al. 2002,
Titarchuk \& Wood 2002).

In  conclusion  it is remarked  that the model described here is
a simple and
rough one. Many physical
details have been  neglected. These include
 NS spin induced gravitomagnetic effect,
NS quadruple induced nodal precession
(Vietri and Stella 1998),  the self-gravity of the disk,
magnetosphere structure and magnetic axis inclination, the spiral-in effect
of the accreted matter, the origination and influence of the non-zero
eccentricity, etc..
 In particular,
the non-zero eccentricity should have some  effects on the
QPOs  but how this would originate   is still unknown.
 We can at least
speculate that the motion of the accreted matter in the disks
might  not be described  exactly  by
a free test particle in a circular orbit of a purely gravitational field.
Consideration of these factors will inform  our future exploration
and understanding  of QPOs. Nevertheless, our
proposed QPO frequencies are all inversely  related to the radii of the
accreted disks, and the instantaneous accretion disk radius is inversely
related to the mass  accretion rate (see e.g., Shapiro \& Teukolsky 1983),
which suggests that  the QPO frequency would
be proportional to  the mass accretion rate.
This conclusion is consistent with the QPO detections of
Z-sources and Atoll sources  as well as BHCs (van der Klis 2000).

\vskip .3cm

\begin{acknowledgements}
Thanks are due to T. Belloni, M. M\'endez and  D. Psaltis for
providing the QPO data, and discussions with   T.P. Li and
W. Zhang are highly appreciated. The author is  thankful to the
anonymous referee for critical and  valuable comments, and
suggestions for  the revision of the paper.

\end{acknowledgements}

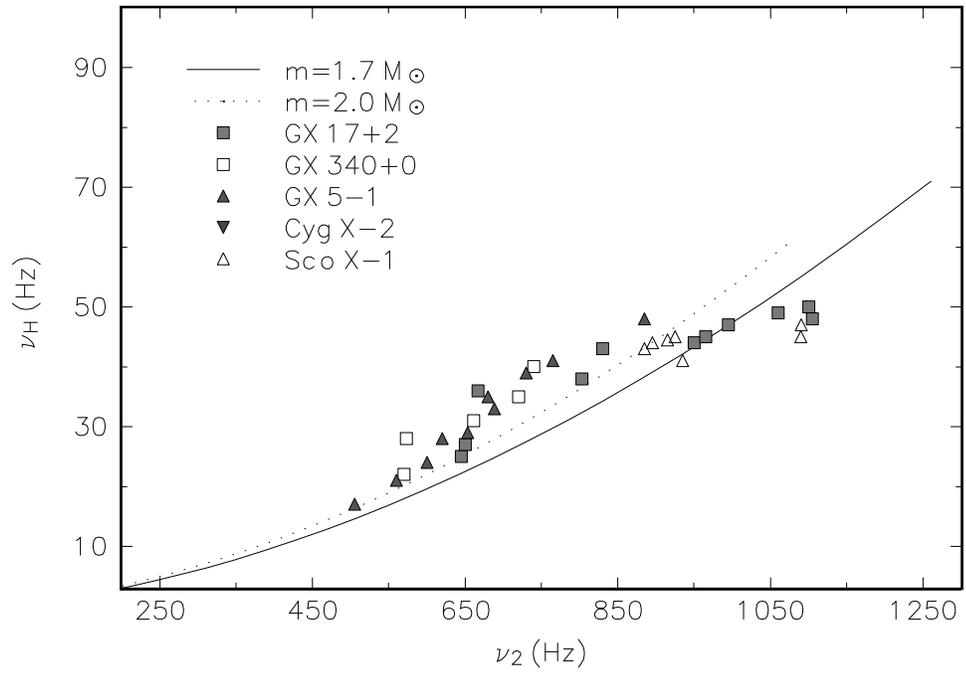
\begin{figure}
\caption[fig1]
{ HBO frequency  versus the upper kHz QPO
frequency  for five Z sources of  LMXBs
(Psaltis et al. 1998, 1999  and references therein).
Error bars are not plotted for the sake of clarity.
The model presents a good  consistency  for  the  nearly circular
orbit of NS mass about 2.0 solar mass with the scaling parameter
$\phi = 0.4$.}
\label{fig1}
\end{figure}

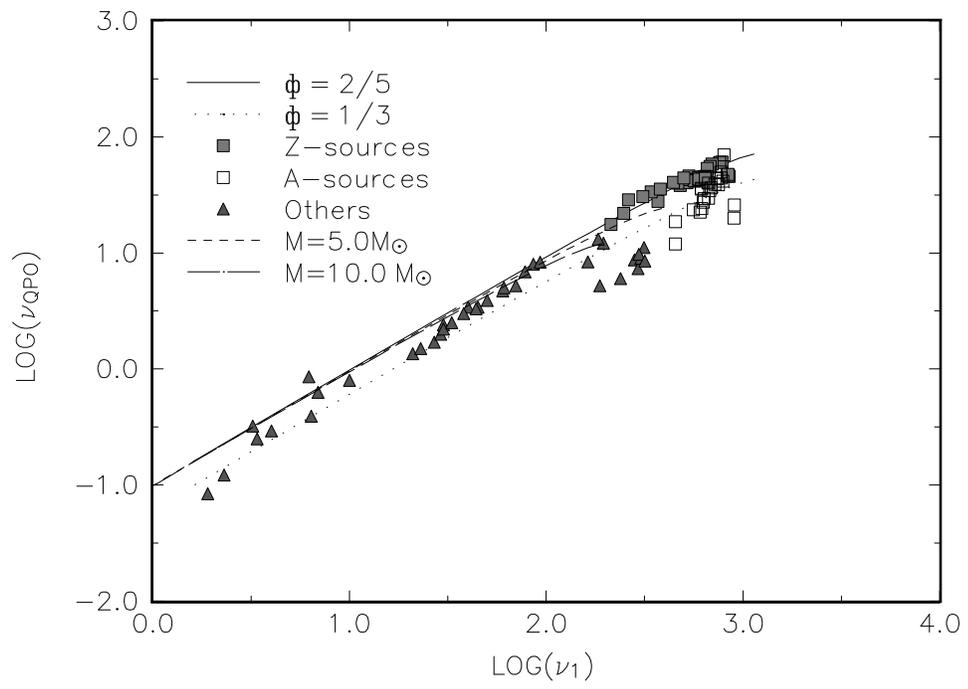
\begin{figure}
\caption[fig2]
{The lower kHz QPO frequency   versus the low QPO
frequency   for the Z-sources, Atoll sources and other sources
(Psaltis et al. 1999). Error
bars are not plotted for the sake of clarity.
The theoretical  curves are only weakly related to the NS mass
parameter.
The solid(dot) line represents the parameter condition  m=2.0
and  $\phi=0.4$ ($\phi = 0.33$);
the short(long) dashed  line represents the parameter condition
m=5.0(10.0) $\ms$ and $\phi=0.4$.}
\label{fig2}
\end{figure}


\begin{thebibliography}{*}

\bibitem{}
Abramowicz, M.~A., \& Kluzniak, W. 2001, A\&A, 374, L19


\bibitem{}
Abramowicz, M.~A., Bulik, T., Bursa, M., \& Kluzniak, W. 2003, A\&A,
404, L21




\bibitem
{}{}Alpar, A.,  \& Shaham, J. 1985, \nat,  {\bf 316}, 239

\bibitem{} Belloni,T.,
Psaltis, D.,  \& van der Klis, M., 2002, \apj, 572, 392.















\bibitem{} Ford, E.C., van der Klis, M., van Paradijs, J.,
 M\'endez, M., Wijnands, R., \& Kaaret, P. 1998,
\apjl, 508, L155



\bibitem{}
Hasinger, G., \& van der Klis, M.
1989, \aap, 225, 79

\bibitem{}
Klein, R.L., Jernigan, G.J., Arons, J., Morgan, E.H., \&
Zhang, W. 1996, \apj, 469, L119

\bibitem{}
Lamb, F.K., Shibazaki, N., Alpar, M.A., \& Shaham, J.
1985, \nat, 317, 681



\bibitem{} M\'endez, M., van der Klis, M., van Paradijs, J.,
Lewin,
 W.H.G.,
Vaughan, B.A., et al. 1998a, \apjl, 494, L65

\bibitem{} M\'endez, M., van der Klis, M., Wijnands, R., Ford,
E.C.,
van Paradijs, J., \& Vaughan, B.A.
1998b, \apjl,  505, L23


\bibitem{}
M\'endez, M., \& van der Klis, M.  1999, \apj , 517, L51


\bibitem{}
Miller, M.\ C., Lamb, F.\ K., \& Psaltis, D.\ 1998, \apj, 508, 791



\bibitem{}
Muno, M.P. 2004,
"Millisecond Oscillations During Thermonuclear X-ray Bursts",
review article for "X-Ray Timing 2003: Rossi and
Beyond", eds. P. Kaaret, F. K. Lamb, \& J. H.
Swank (Melville, NY: American Institute of Physics)

\bibitem{}
Remillard, R. A., Muno, M. P., McClintock, J. E.,  \& Orosz, J. A.
2002, \apj, 580, 1030



\bibitem{} Orosz, J.A., \& Kuulkers,  E. 1999, \mnras, 305, 1320
(astro-ph/9901177)

\bibitem{} Osherovich, V., \&
Titarchuk, L. 1999, ApJ, 522, L113





\bibitem{}Psaltis, D., Mendez, M.,
Wijnands, R., Homan, J., Jonker, P., van der Klis,
M., Lamb, F.\,K., Kuulkers, E., van Paradijs, J., \&
Lewin, W.\,H.\,G.\ 1998, \apj, 501, L95




\bibitem{}
Psaltis, D., Belloni, T., \& van der Klis, M.\ 1999, \apj,
 520, 262  (astro-ph/9902130)

\bibitem{}Psaltis, D., 2000,
   Advances for Space Research, 481, 281; invited review talk at the 33rd COSPAR Scientific Assembly, Warsaw, Poland,
astro-ph/0012251



\bibitem{} Psaltis, D., \& Norman, C. 2000, ApJ,
(astro-ph/0001391)

\bibitem{} Shapiro, S.L.,  \& Teukolsky, S.A. 1983, {\it Black Holes,
 White Dwarfs and Neutron Stars} (Wiley, New York)






\bibitem{}Stella, L., \& Vietri, M., 1999,  \prl,  82, 17 (SV99)

\bibitem{}Strohmayer, T., \&  Bildsten, L., 2003,
 To appear in Compact Stellar X-Ray Sources,
eds. W.H.G. Lewin and M. van der Klis, Cambridge
University   Press: Astro-ph/0301544


\bibitem{}
Strohmayer, T., Zhang, W., Smale, A., Day, C., Swank, J.,
Titarchuk, L., \& Lee, U. 1996, \apjl, 469, L9



\bibitem{}
Titarchuk, L., Lapidus, I.I., \& Muslimov, A. 1998, \apj, 315, 499


\bibitem{}
Titarchuk, L., \& Wood, K. 2002, \apj, 577, L23
(astro-ph/0208212)



\bibitem[Wagoner 1999]{W1999}
         Wagoner, R.\,W.\ 1999, Phys.\ Rep., 311, 259



\bibitem{}
van der Klis, M., Wijnands, A.D., Horne, K., \& Chen, W. 1997,
\apj , 481, L97


\bibitem{}
van der Klis, M. 2000, ARA\&A, 38, 717 (astro-ph/0001167)



\bibitem{} Vietri, M., \& Stella, L.\ 1998, \apj, 503, 350












\end{thebibliography}
\end{document}